\documentclass[letterpaper,english,aps,prl,superscriptaddress,reprint]{revtex4-2}
\usepackage[T1]{fontenc}
\usepackage[utf8]{inputenc}
\usepackage{color}
\usepackage{babel}
\usepackage{amstext}
\usepackage{amssymb}
\usepackage{graphicx}
\usepackage{microtype}
\usepackage[unicode=true,
 bookmarks=true,bookmarksnumbered=false,bookmarksopen=false,
 breaklinks=true,pdfborder={0 0 0},pdfborderstyle={},backref=false,colorlinks=true]
 {hyperref}
\hypersetup{pdftitle={Long-radial-range rorrelation turbulence},
 pdfauthor={Rongjie Hong},
 allcolors=blue}

\makeatletter

\pdfpageheight\paperheight
\pdfpagewidth\paperwidth

\usepackage{newtxtext}
\usepackage[varg]{newtxmath}

\makeatother

\begin{document}
\title{Observation of Long-Radial-Range-Correlation in Turbulence in High-Collisionality
High-Confinement Fusion Plasmas}
\author{R.~Hong}
\affiliation{Physics and Astronomy Department, University of California, Los Angeles,
CA 90095, USA}
\author{T.~L.~Rhodes}
\affiliation{Physics and Astronomy Department, University of California, Los Angeles,
CA 90095, USA}
\author{P.~H.~Diamond}
\affiliation{Center for Astrophysics and Space Sciences, University of California,
San Diego, La Jolla, CA 92093, USA}
\author{Y.\,Ren}
\affiliation{Princeton Plasma Physics Laboratory, Princeton, NJ 08543, USA}
\author{L.~Zeng}
\affiliation{Physics and Astronomy Department, University of California, Los Angeles,
CA 90095, USA}
\author{X.~Jian}
\affiliation{Center for Energy Research, University of California, San Diego, La
Jolla, CA 92093, USA}
\author{K.~Barada}
\affiliation{Physics and Astronomy Department, University of California, Los Angeles,
CA 90095, USA}
\author{G.~Wang}
\affiliation{Physics and Astronomy Department, University of California, Los Angeles,
CA 90095, USA}
\author{W.~A.~Peebles}
\affiliation{Physics and Astronomy Department, University of California, Los Angeles,
CA 90095, USA}
\begin{abstract}
We report on the observation of spatially asymmetric turbulent structures
with a long radial correlation length in the core of high-collisionality
\emph{H}-mode plasmas on DIII-D tokamak. These turbulent structures
develop from shorter wavelength turbulence and have a radially elongated
structure. The envelope of turbulence spans a broad radial range in
the mid-radius region, leading to streamer-like transport events.
The underlying turbulence is featured by intermittency, long-term
memory effect, and the characteristic spectrum of self-organized criticality.
The amplitude and the radial scale increase substantially when the
shearing rate of the mean flow is reduced below the turbulent scattering
rate. The enhanced LRRC transport events are accompanied by apparent
degradation of normalized energy confinement time. These findings
constitute the first experimental observation of long-radial-range
turbulent transport events in high-collisionality \emph{H}-mode plasmas,
and demonstrate the role of mean shear flows in the regulation of
turbulence with long-radial-range correlation.
\end{abstract}
\maketitle
The classic approach of describing transport and relaxation adopts
the Chapman-Enskog theory and presumes a clear separation of scales
between fluctuations and macroscopic systems. However, long-range
correlation can develop and act to drive scale-invariant evolution.
The long-range correlation events can be intermittent, as in avalanches
in self-organized criticality, or coherent, as in large-eddy circulation.
Such transport events have been observed in incompressible fluids
\citep{hof2004experimental}, plasmas \citep{yamada2008anatomy},
neural activities \citep{bedard2006doesthe}, complex networks \citep{albert2002statistical},
phase transitions \citep{hung2011observation}, etc. In any of those
cases, the key issues are the characterization of long-range events
and the understanding of their origins. 

The long-range correlation is also highly relevant to studies of magnetic
fusion energy. One major challenge for fusion plasma physics is to
predict and control turbulence and transport. The conventional turbulent
transport model is based on a set of local expressions of flux-gradient
relations \citep{doyle2007chapter}. However, there are several indications
that the local model is inadequate \citep{ida2015towards}. One of
the most prominent examples is the breakdown of gyro-Bohm scaling
that is a fundamental element of the local approach \citep{doyle2007chapter,hahm2018mesoscopic}.
Such a departure implies that the radial scales of turbulent structures
in fusion plasmas are likely in excess of the expected turbulence
cell size. That is, turbulent structures with a long-radial-range
correlation (LRRC) may exist in fusion plasmas. 

In fusion plasmas, turbulent structures with LRRC have been predicted
by theory and simulations, including avalanches \citep{diamond1995onthe},
turbulence spreading \citep{hahm2004turbulence}, streamers \citep{dorland2000electron},
etc. There are some experimental observations for such structures
from low-confinement mode (\emph{L}-mode) discharges \citep{gentle1995strongnonlocal,nazikian2005measurement,hamada2006streamers,inagaki2011observation,ji2016onthe}.
However, until now, it remains an open question whether such LRRC
mode can occur in the high-confinement mode (\emph{H}-mode), which
is the preferred operation scenario for ITER and fusion reactors.
In addition, substantial degradation of plasma confinement in the
\emph{H}-mode discharges has been widely observed when the collisionality
is raised \citep{luce2008application} or the averaged density approaches
the Greenwald density \citep{doyle2007chapter}. Further analyses
found that, as collisionality is increased, the core electron thermal
diffusivity increases substantially, while the ion thermal diffusivity
is similar \citep{petty1999scaling,kaye2007confinement,valovivc2011collisionality}.
It is an indication that core turbulent transport in the electron
channel, due to an increase in intensity or the radial scale, plays
an important role in the confinement degradation. Therefore, it is
of practical importance and interest to identify and characterize
core turbulence and their correlations that emerge in high-collisionality
or high-density \emph{H}-mode fusion plasmas. 

In this letter, we show that shorter wavelength turbulence develops
into spatially asymmetric turbulent structures with a long-radial-range
correlation in the mid-radius region of high-collisionality \emph{H}-mode
plasmas on the DIII-D tokamak. The magnitude and the radial scale
of those turbulent structures increase significantly when the $E_{r}\times B$
mean flow shearing rate decreases. These findings thus provide the
first experimental evidence for the existence of the LRRC turbulent
structures and resulting transport events in \emph{H}-mode plasmas
at high collisionality, as well as the potential role of the radially
sheared $E_{r}\times B$ mean flow in regulating the long-range transport
dynamics. The emergence of such LRRC transport events may serve as
a candidate explanation for the degrading nature of \emph{H}-mode
core plasma confinement at high-collisionality.

The experiments were performed using upper single-null plasmas with
closed divertors on the DIII-D tokamak \citep{luxon2002adesign}.
A dimensionless collisionality scan was performed by varying the toroidal
magnetic field $B_{t}$ and the plasma current $I_{p}$ by a factor
of $1.6$ with fixed $B_{t}/I_{p}$, which is a standard approach
used in previous dimensionless scan experiments on DIII-D \citep{petty1999scaling}.
In these discharges, the collisionality ($\nu_{e}^{*}\sim n_{e}/T_{e}^{2}$)
is scanned, with well-matched plasma density, $n_{e}$, and some transport-relevant
dimensionless parameters in the mid-radius region, e.g., $\rho^{*}$
(ratio of ion gyro-radius to plasma minor radius $\rho_{i}/a$), $\beta_{N}=\frac{\beta}{I/aB_{t}}$
(normalized ratio of plasma to magnetic pressure), $q$ (safety factor),
$T_{e}/T_{i}$ (electron-to-ion temperature ratio), $\kappa$ (elongation),
$\delta$ (triangularity). The neutral beam injection of $P_{\mathrm{NBI}}=7$
MW provides constant input torque and heating power in each shot.
Large-scale magnetohydrodynamics (MHD) instabilities are mitigated
and of similar amplitude ($\tilde{B}_{\theta}<0.1$ mT ) in these
shots. The line-averaged plasma density is $\bar{n}_{e}=4-4.2\times10^{19}\,\mathrm{m}^{-3}$
in these discharges. The corresponding Greenwald fraction is raised
from about 0.5 to 0.9. Note that the pedestal conditions and edge
localized modes (ELMs) characteristics changed dramatically during
the collisionality scan. To avoid confusion, the analysis presented
in this study is focused on the profiles and fluctuations between
ELMs. The electron density and temperature profiles are measured using
Thomson scattering diagnostics \citep{eldon2012initial}. The profiles
of carbon ion density, temperature, and velocities are measured by
the charge exchange recombination (CER) diagnostic system \citep{chrystal2016improved}.
These carbon ion profiles are used to infer the mean $E_{r}\times B$
shear flow using the ion force balance equation \citep{burrell1997effects}.
The multi-channel Doppler backscattering (DBS) diagnostics \citep{peebles2010anovel}
are used to detect the electron density fluctuations at the wavenumber
range of $5<k_{\perp}<10\,\mathrm{cm^{-1}}$ ($1<k_{\perp}\rho_{s}<4$
where $\text{\ensuremath{\rho_{s}}}$ is the ion gyro-radius with
sound speed) in the mid-radius region ($0.35<\rho<0.8$ with $\rho$
the normalized minor radius).

\begin{figure}
\includegraphics[width=0.95\columnwidth]{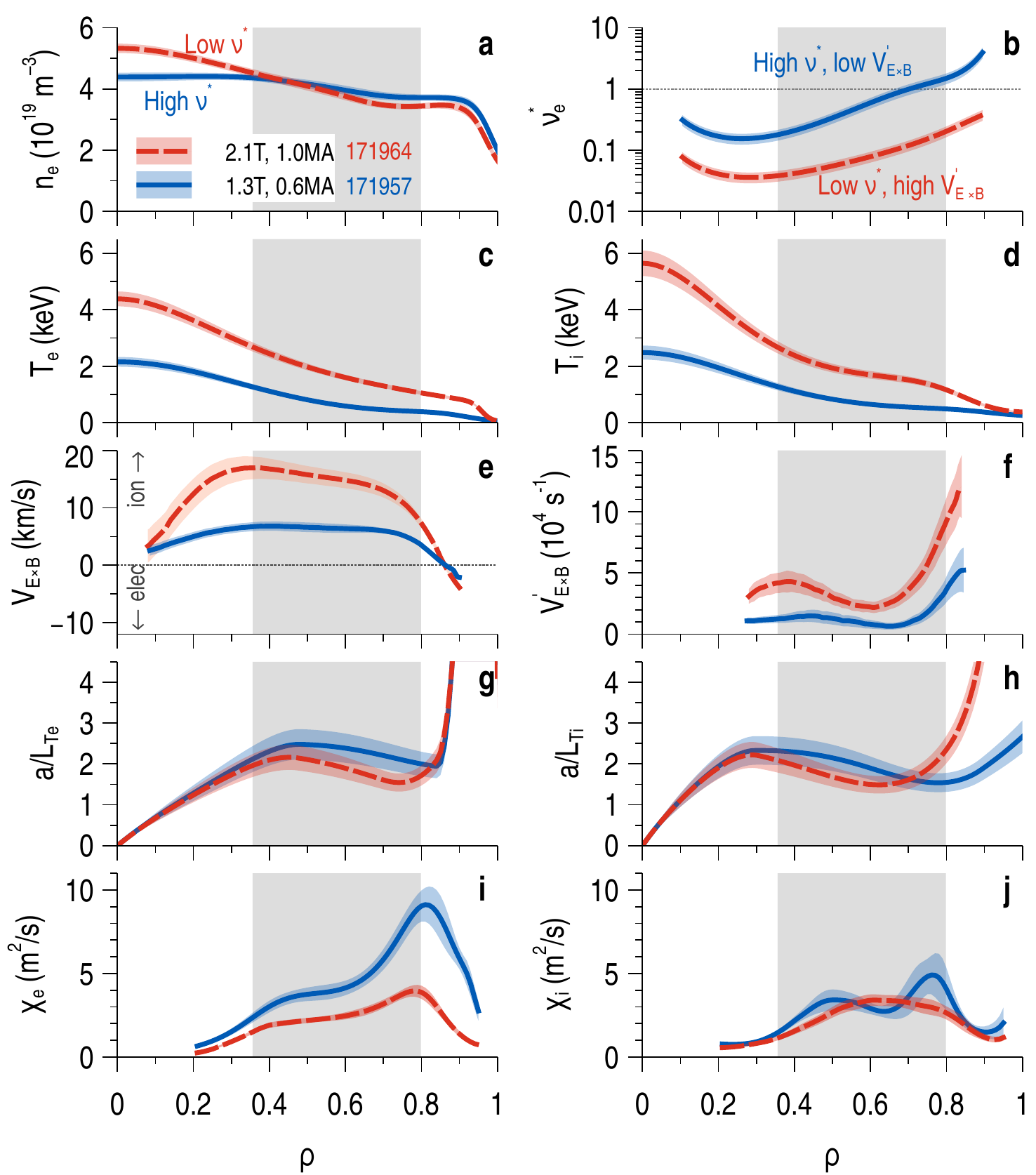}

\caption{\label{fig:profs} Radial profiles of (a) the electron density, (b)
the electron collisionality, (c) the electron temperature, (c) the
ion temperature, (e) the mean $E_{r}\times B$ shear flow, (f) the
mean flow shearing rate, (g) the normalized electron temperature gradient,
(h) the normalized ion temperature gradient, (i) effective electron
thermal diffusivity, and (j) effective ion thermal diffusivity. The
blue curves represent profiles from a high collisionality discharge,
and the red curves corresponds to those in a low collisionality shot.
Gray shadings indicate the region of interest in this study.}
\end{figure}

Figure \ref{fig:profs} shows the relevant radial profiles at two
distinct values of the collisionality. The low collisionality discharge
has a more peaked density profile (Figure \ref{fig:profs}(a)), consistent
with previous DIII-D observations \citep{mordijck2020collisionality}.
The electron and ion temperature profiles are varied with $T_{e,i}\propto B_{t}^{2}$
(Figures \ref{fig:profs}(c) and (d)), such that the $T_{e}/T_{i}$
ratio and the gyro-radii $\text{\ensuremath{\rho_{s}}}$ are kept
nearly constant in the mid-radius region during the scan. The corresponding
electron collisionality, $\nu_{e}^{*}$, is increased by a factor
of 7 in the mid-radius region (Figure \ref{fig:profs}(b)). The normalized
electron and ion temperature gradient, $L_{T}^{-1}=-\nabla_{r}\ln T$,
are similar during the scan (Figures \ref{fig:profs}(g) and (h)).
The toroidal rotation drops when the collisionality is raised. As
a result, the mean $E_{r}\times B$ shear flow and its shearing rate
reduced substantially at higher collisionality (Figures \ref{fig:profs}(e)
and (f)). Here, the $E_{r}\times B$ mean flow shearing rate , $V_{E\times B}^{\prime}$,
is calculated using the Waltz-Miller formulation \citep{waltz1999iontemperature}.
Moreover, the core thermal diffusivity, given by the local power balance
analysis using the \textsc{transp} code, increases by a factor of
2--3 at high collisionality (Figuers \ref{fig:profs}(i)). The effective
ion thermal diffusivity is only marginally higher in high collisionality
shots (Figuers \ref{fig:profs}(j)).

\begin{figure}
\includegraphics[width=1\columnwidth]{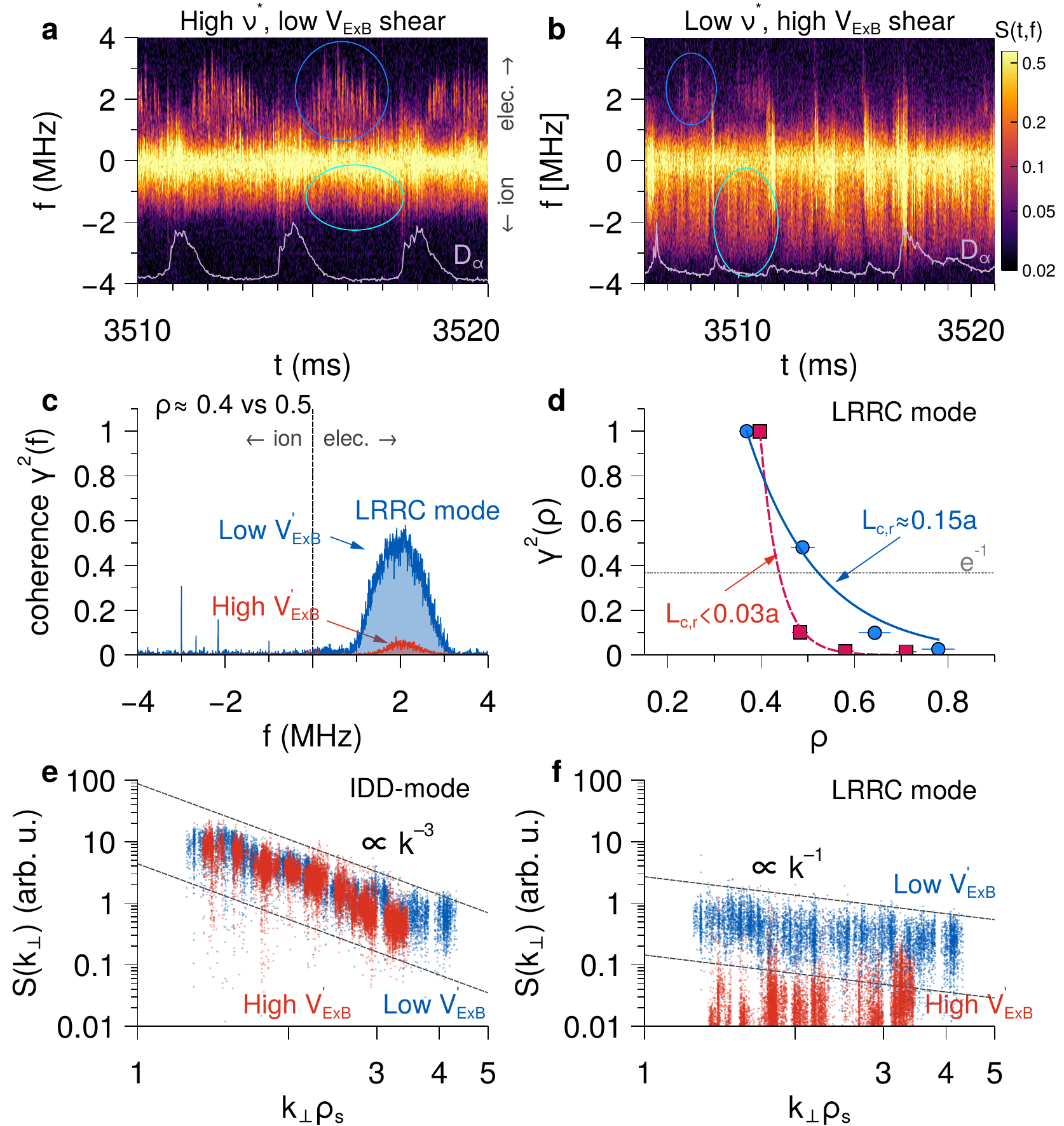}

\caption{\label{fig:specs}First row: spectrogram of density fluctuations measured
by the DBS diagnostics at $\rho\approx0.5$ in high collisionality
(a) and low collisionality (b) discharges, respectively. Here, the
ELM evolutions are indicated by the $D_{\alpha}$ emission intensity
(gray curves). Second row: (c) Coherence between local density fluctuations
at $\rho\approx0.4$ and $0.5$; (d) Radial profiles of the peak cross-coherence
for the LRRC turbulence, where the channel at $\rho_{0}\approx0.4$
is chosen as the reference. The curves represent the exponential fits,
i.e., $\gamma^{2}(\rho)=\exp(-\frac{\rho-\rho_{0}}{L_{c,r}})$, where
$L_{c,r}$ represents the radial correlation length in terms of the
minor radius.. Third row: the wavenumber power spectra of IDD-mode
(e) and LRRC mode (f), respectively.}
\end{figure}

With different levels of collisionality and $E_{r}\times B$ shear
flows, density fluctuations exhibit differing spectral and temporal
characteristics. In the high collisionality discharges, with a weak
$E_{r}\times B$ shear flow, two different modes can be detected by
the DBS diagnostics between ELMs (Figure \ref{fig:specs}(a)). Here,
the ELM evolutions are indicated by the $D_{\alpha}$ emission intensity
(gray curves). One mode has a negative Doppler frequency shift corresponding
to ion diamagnetic drift direction (IDD) in the lab frame (IDD-mode
marked by cyan circles); the other mode has a positive Doppler frequency
shift around 2 MHz propagating in the electron diamagnetic drift direction
(EDD) in the lab frame (EDD-mode marked by blue circles). In low collisionality
discharges, with strong $E_{r}\times B$ shear flow, the IDD-mode
with $f<0$ can still be observed with a larger Doppler shift, but
the EDD-mode around $f\approx2$ MHz is weak and rarely happened (Figure
\ref{fig:specs}(b)). Note that the scales of both the IDD-mode and
EDD-mode are in the sub-ion-gyroradius regime.

In this study, the EDD-mode is of interest as it exhibits a long-radial-range
correlation, i.e., a LRRC mode. Figure \ref{fig:specs}(c) shows the
frequency resolved coherence $\gamma^{2}(f)$ between electron density
fluctuations at $\rho\approx0.4$ and 0.5. The fluctuations at $f<0$
do not show any clear coherence between different radial locations,
while the LRRC mode at $f\approx2$ MHz shows substantial coherence
with lower mean $E_{r}\times B$ shear flow. Figure \ref{fig:specs}(d)
shows the radial profiles of the peak coherence for underlying turbulence
of the LRRC mode, using the measurements at $\rho\approx0.4$ as the
reference. The radial correlation length can be obtained via the decay
length of the exponential fits, i.e., $\gamma^{2}(\rho)=\gamma^{2}(\rho)=\exp(-\frac{\rho-\rho_{0}}{L_{c,r}})$,
where $L_{c,r}$ represents the radial correlation length in terms
of the minor radius fraction. The radial correlation length of the
LRRC mode, $L_{c,r}$, is no more than 2 cm (plasma minor radius $a\approx61$
cm) with strong $E_{r}\times B$ shear flow, while it increases to
about 10 cm with reduced $E_{r}\times B$ shear flow in high collisionality
plasmas. This long radial correlation length indicates a radially
elongated structure of the LRRC mode, i.e., $k_{r}\ll k_{\theta}$
with $k_{r}\rho_{s}=0.1-0.3$ and $k_{\theta}\rho_{s}=1-4$.

The wavenumber power spectra of electron density fluctuations are
displayed in Figures \ref{fig:specs}(e) and (f). It is found that
the wavenumber power spectra of the IDD-mode obey a power-law of $S(k_{\perp})\propto k_{\perp}^{-3}$
and are similar in the magnitude for discharges with low and high
$E_{r}\times B$ shear flows (Figure \ref{fig:specs}(e)). The negligible
variations in magnitude indicate that it is \emph{not} likely responsible
for the changes in the plasma confinement during the collisionality
scan. On the other hand, the power spectrum of LRRC mode is much larger
in the weak $E_{r}\times B$ shear and high collisionality discharges,
and shows a power-law of $S(k_{\perp})\propto k_{\perp}^{-1}$ (Figure
\ref{fig:specs}(f)). The $1/k$ spectrum of the LRRC mode is indicative
of its avalanching-like behavior that are commonly associated with
self-organized criticality (SOC) \citep{bak1987selforganized,diamond1995onthe}.

\begin{figure}
\includegraphics[width=0.95\columnwidth]{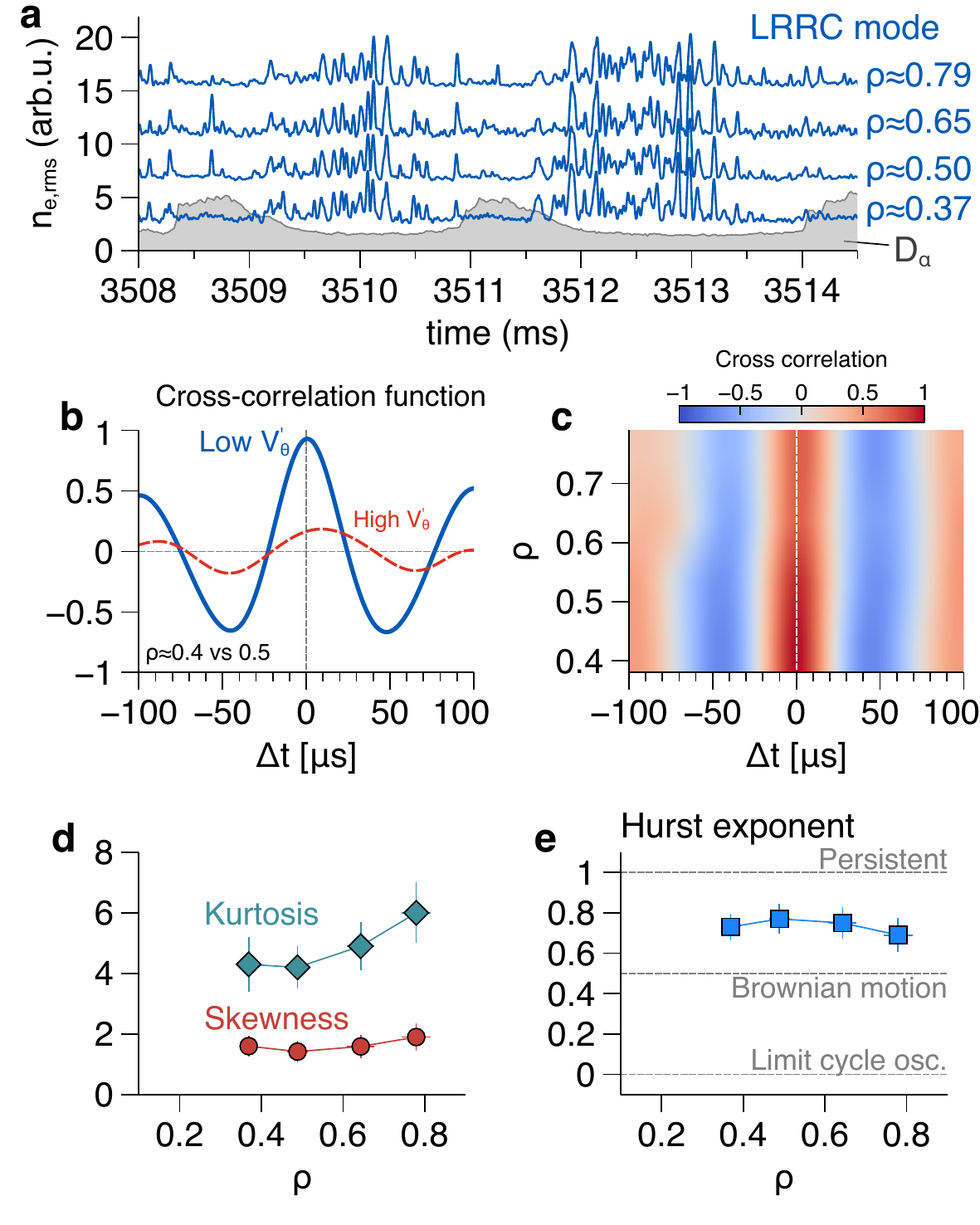}

\caption{\label{fig:stats} (a) Time evolution of the envelopes (RMS levels)
of local density fluctuations (with offsets) corresponding to the
LRRC mode at different radial locations. (b) Cross-correlation function
between RMS levels at $\rho\approx0.4$ and 0.5 for two different
cases. (c) Contours of the cross correlation function between RMS
levels for each DBS channel, with inner most channel as the reference.
(d) Skewness and Kurtosis and (e) the Hurst exponent of the underlying
fluctuations of LRRC turbulence. The error bars indicates the standard
error of the statistical analysis.}
\end{figure}

The envelopes of the LRRC mode also show a clear cross-correlation
across a wide radial range. As shown in Figure \ref{fig:stats}(a),
the envelopes, or root-mean-square (RMS) levels, of density fluctuations
corresponding to the LRRC mode at different radial locations are well
aligned in time. Correlation analysis reveals that the cross-correlation
coefficient between the envelopes at $\rho\approx0.4$ and 0.5 approaches
the value of one with weak flow shear, while the one in strong flow
shear case is insignificant (Figure \ref{fig:stats}(b)). Note that
the amplitude of the LRRC mode reduced substantially in the strong
flow case. This value is well below the By plotting the cross-correlation
of the envelopes as a function of the radial location and the time
lag, for the weak flow case (Figure \ref{fig:stats}(c)), with inner
most channel as the reference, one can see that the envelope of the
LRRC mode spans a broad radial range of spatial scales ($\rho_{i}\ll\Delta_{r}^{\mathrm{env}}\lesssim a$)
with a negligible time delay (no more than a few micro-seconds). Such
a wide radial scale of the envelopes of shorter wavelength turbulence
indicates the emergence of streamer-like transport events in the weak
flow shear discharges.

The statistical analysis of the underlying density fluctuations corresponding
to the LRRC mode show large values of skewness ($S=1-2$) and kurtosis
($K=4-6$) (Figure \ref{fig:stats}(d)), indicative of highly intermittent
features of underlying turbulence. The Hurst exponent has been calculated
using the technique of rescaled range analysis \citep{gilmore2002investigation},
and is found to range from 0.7 to 0.8 (Figure \ref{fig:stats}(e)).
This Hurst exponent is well above the value of 0.5 that corresponds
to the Brownian motion, indicating a long-term memory characteristic
in the LRRC mode. Here, the Hurst exponent and higher order moments
are calculated for each inter-ELM period, and a total time interval
of 200 ms (about 60 samples) is used to estimate the means and standard
errors.

\begin{figure}
\includegraphics[width=1\columnwidth]{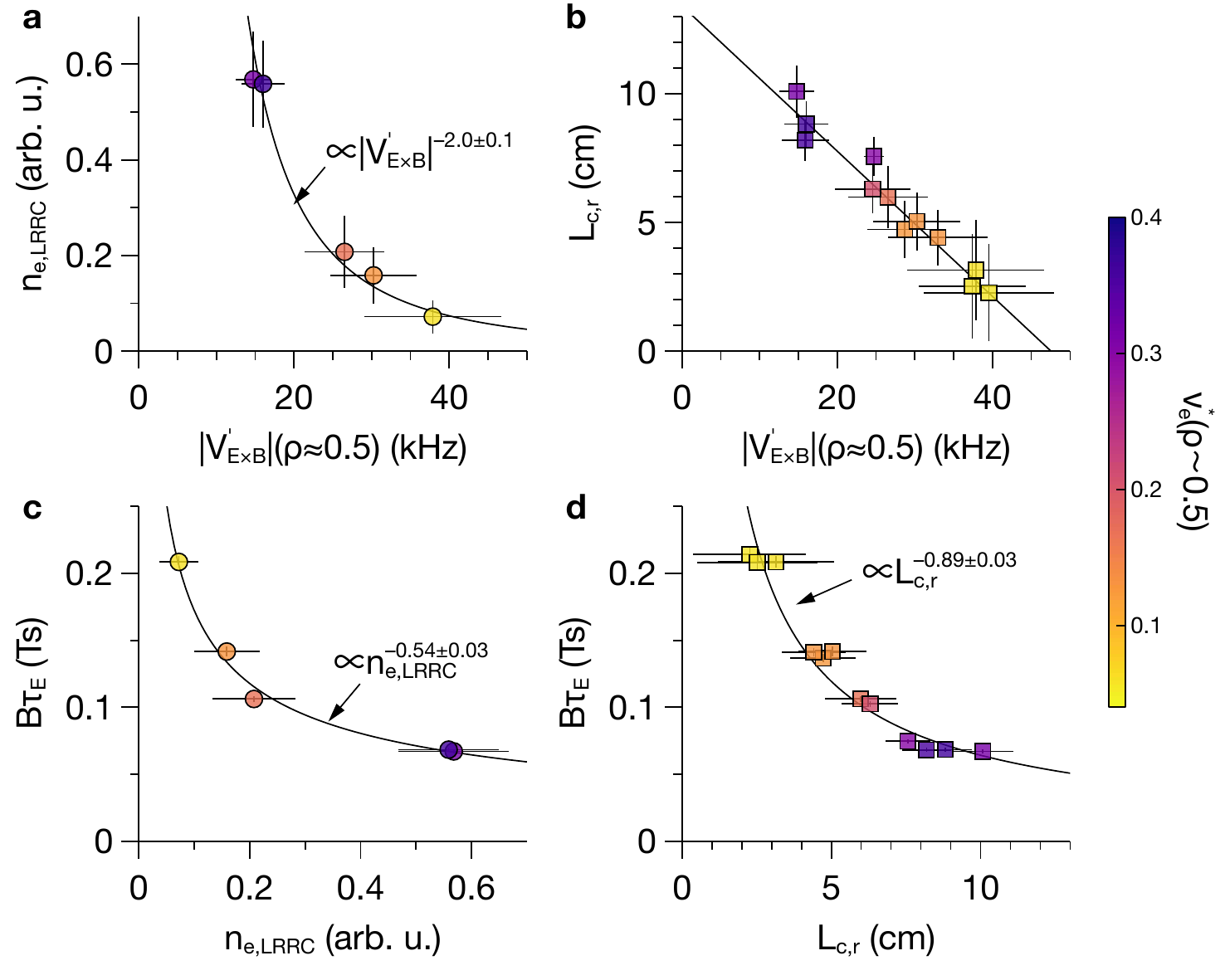}

\caption{\label{fig:scaling} The amplitude (a) and the radial correlation
length (b) of the underlying fluctuations of LRRC transport events
are plotted against the local $E_{r}\times B$ velocity shearing rate
at $\rho\approx0.5$. Normalized confinement time is compared against
the amplitude RMS levels (c) and the radial correlation length of
LRRC turbulence (d), respectively. The collisionality of each discharge
at $\rho\approx0.5$ is color-coded, with brighter markers corresponding
to lower collisionality.}
\end{figure}

To gain further insights into the influence of the $E_{r}\times B$
shear flow on those LRRC turbulent structures, we compare the amplitude
($k_{\perp}\rho_{s}\sim3$) and the radial correlation length of underlying
turbulence of LRRC transport events against the local $E_{r}\times B$
shearing rate at $\rho\approx0.5$ (Figure \ref{fig:scaling}(a)).
The collisionality (at $\rho\approx0.5$) of each discharge is color-coded,
with brighter markers corresponding to lower collisionality discharges.
As shown in Figure \ref{fig:scaling}(a), when the local $E_{r}\times B$
flow shearing rate is increased, the amplitude of the LRRC mode decreases
as $\tilde{n}_{e,\mathrm{LRRC}}\propto|V_{E\times B}^{\prime}|^{-2}$.
The radial correlation length is inversely correlated to the local
$E_{r}\times B$ shearing rate (Figure \ref{fig:scaling}(b)). For
weak flow shear case, the turbulent scattering rate is estimated using
the inverse of the eddy turnover time of LRRC mode, i.e., $\tau_{c}^{-1}\approx20-30\,\mathrm{kHz}$.
Here, the eddy turnover time of LRRC mode is calculated using the
auto-correlation time of the corresponding RMS levels at high collisionality
and weak flow shear, and thus it is independent of Doppler shifts
due to the background mean flow. It is found that the amplitude and
radial scale of the LRRC mode increase substantially when $|V_{E\times B}^{\prime}|<\tau_{c}^{-1}$
(Figure \ref{fig:scaling}(a) and (b)). This finding is an indication
that the mean $E_{r}\times B$ shear flow impacts the development
of the LRRC turbulent structures, consistent with the shear decorrelation
mechanism. It is worth noting that the reduced mean shear layer is
associated with the high collisionality in this study, and further
experiments using net-zero torque input are desirable to distinguish
effects of collisionality and mean shear flows.

The development of LRRC turbulent structures and the resulting transport
events is also associated with degradation of the normalized confinement
time (Figure \ref{fig:scaling}(c) and (d)), which is in conformity
with previous results that the plasma transport increases with collisionality
in \emph{H}-mode plasmas \citep{petty1999scaling}. In particular,
the dependence on the LRRC mode amplitude, $B\tau_{E}\sim\left|\tilde{n}_{e,\mathrm{LRRC}}\right|^{-0.54\pm0.03}$,
appears to be close to the collisionality scaling, $B\tau_{E}\sim\nu_{*}^{-0.56\pm0.06}$,
previously reported on DIII-D \citep{luce2008application}. Note that
enhanced turbulent transport here is not likely driven by ion-scale
long-wavelength turbulence ($k_{\theta}\rho_{s}<1$) which is reported
to slightly decrease in the core when collisionality is raised by
a factor of 5 \citep{mordijck2020collisionality}. Also, nonlinear
gyro-kinetic simulations show that the calculated ion-scale turbulent
transport underestimates the total electron heat flux by an order
of magnitude, particularly in high collisionality plasmas \citep{holland2017gyrokinetic},
implying the essential role of the higher-$k_{\theta}$ turbulence
($k_{\theta}\rho_{s}>1$) in electron transport processes at high
collisionality. These findings suggest that the emergence of LRRC
transport events may serve as a candidate explanation for confinement
degradation in DIII-D \emph{H}-mode fusion plasmas at higher collisionality
\citep{petty1999scaling,luce2008application}.

\begin{figure}
\includegraphics[width=1\columnwidth]{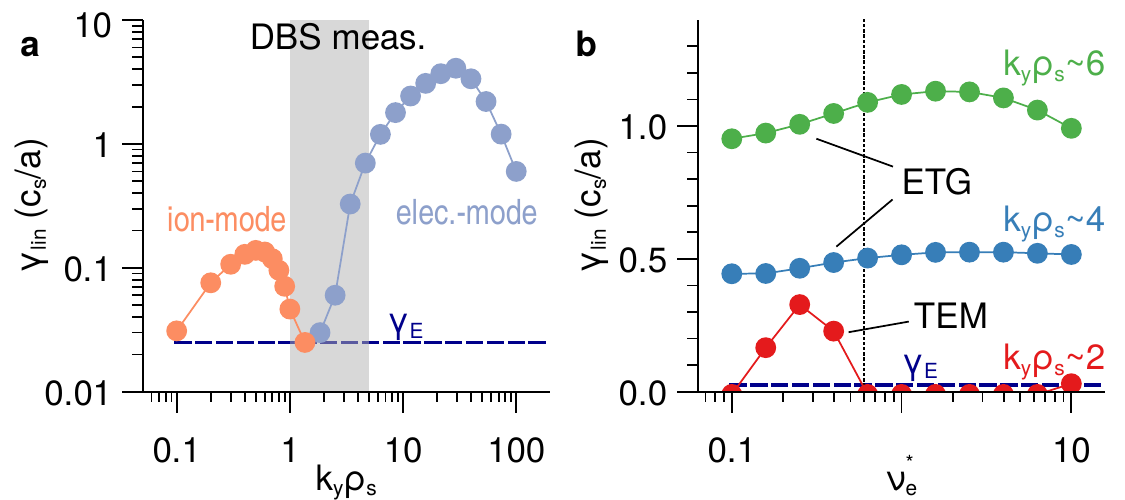}

\caption{\label{fig:Linear}Linear growth rate calculation using \textsc{CGYRO}
code for (a) wavenumber scan in high collisionality plasma at $\rho\approx0.6$
and (b) collisionality scan at three different wavenumbers. The black
dashed line in (b) indicates the experiment value of local collisionality.}

\end{figure}

Linear gyrokinetic simulations have been performed to identify the
underlying turbulence of those LRRC transport events using the\textsc{
cgyro} code. It is found that electrostatic instabilities with $k_{\theta}\rho_{s}>2$
are linearly unstable (Figure \ref{fig:Linear}(a)). Two common instabilities
in this range are the trapped electron mode (TEM) and the electron
temperature gradient mode (ETG). The TEM is found to be linearly stabilized
in high collisionality plasmas in this study, via a collisionality
scan (red in Figure \ref{fig:Linear}(b)), so it is unlikely responsible
for the long-radial-range-correlation transport events. The ETG mode,
on the other hand, is still linearly unstable at higher collisionality
according to the gyrokinetic simulation (green and blue Figure \ref{fig:Linear}(b)).
Previous nonlinear gyrokinetic simulations suggest ETG mode can lead
to streamer-like transport events \citep{dorland2000electron,jenko2002prediction,holland2017gyrokinetic},
which seems to agree with the observations in this study.

In summary, the shorter wavelength ($1<k_{\theta}\rho_{s}<4$) core
turbulence can develop into long-radial-range correlated turbulent
structures in high-collisionality \emph{H}-mode plasmas with reduced
mean $V_{E\times B}$ shear flow on the DIII-D tokamak. These turbulent
structures are radially elongated, and their envelopes span a wide
range in the mid-radius region, leading to streamer-like transport
events. The underlying turbulence shows statistical features that
are usually associated with self-organized criticality, including
intermittency (large skewness and kurtosis), long-term memory effect
(large Hurst exponent), and characteristic power spectrum ($S(k_{\perp})\propto k^{-1}$).
The amplitude and the radial scale of the turbulent structures increase
substantially once the mean flow shearing rate is decreased below
the turbulent scattering rate. The observations summarized here constitute
the first experimental demonstration of the nonlocal turbulent transport
events in high-confinement fusion plasmas, and also provide evidence
for the role of $V_{E\times B}$ shear flow in regulating the long-range
correlated turbulent structures.

The authors greatly appreciate the effort and support of the entire
DIII-D team in performing these experiments. We would like to acknowledge
valuable discussions with M.~E.~Austin, N.~A.~Crocker, and G.~R.~McKee.
This material is based upon work supported by the U.S. Department
of Energy, Office of Science, Office of Fusion Energy Sciences, using
the DIII-D National Fusion Facility, a DOE Office of Science user
facility, under Awards DE-FC02-04ER54698 and DE-SC0019352. 

Disclaimer: This report was prepared as an account of work sponsored
by an agency of the United States Government. Neither the United States
Government nor any agency thereof, nor any of their employees, makes
any warranty, express or implied, or assumes any legal liability or
responsibility for the accuracy, completeness, or usefulness of any
information, apparatus, product, or process disclosed, or represents
that its use would not infringe privately owned rights. Reference
herein to any specific commercial product, process, or service by
trade name, trademark, manufacturer, or otherwise does not necessarily
constitute or imply its endorsement, recommendation, or favoring by
the United States Government or any agency thereof. The views and
opinions of authors expressed herein do not necessarily state or reflect
those of the United States Government or any agency thereof.

\bibliographystyle{apsrev4-2}
\bibliography{LRC_manuscript}

\end{document}